\documentclass[amssymb,amsmath,aps,showpacs,floatfix,nofootinbib,showpacs,12pt]{article}
\pdfoutput=1
\usepackage{latexsym}
\usepackage{color}
\usepackage{verbatim}
\usepackage[section]{placeins}
\usepackage{graphicx}
\usepackage{amsfonts}
\usepackage{amsmath}
\usepackage{amssymb}
\usepackage{booktabs}
\usepackage{longtable}
\usepackage{indentfirst}
\usepackage[bookmarks=false,linktocpage,colorlinks,linkcolor=red,citecolor=blue,urlcolor=magenta]{hyperref}
\usepackage[all]{hypcap}
\begin{document}
\title{\bf Cosmic Rays during BBN as Origin of Lithium Problem}
\author{Ming-ming Kang $^1$, Yang Hu $^2$\thanks{Corresponding author: younger@pku.edu.cn}, Hong-bo Hu $^1$ and Shou-hua Zhu $^{2,3}$}

\maketitle

\begin{center}
{\em $^1$ Key Laboratory of Particle Astrophysics, Institute of High Energy Physics, Chinese Academy of Sciences, Beijing 100049, P. R. China\\
$^2$ Institute of Theoretical Physics $\&$ State Key Laboratory of Nuclear Physics and Technology, Peking University, Beijing 100871, P.R. China\\
$^3$ Center for High Energy Physics, Peking University, Beijing 100871, P.R. China}
\end{center}

\begin{abstract}

There may be non-thermal cosmic rays during big-bang nucleosynthesis (BBN) epoch (dubbed as BBNCRs). This paper investigated whether such BBNCRs can be the origin of Lithium problem or not. It can be expected that BBNCRs flux will be small in order to keep the success of standard BBN (SBBN). With favorable assumptions on the BBNCR spectrum between 0.09 -- 4 MeV, our numerical calculation showed that extra contributions from BBNCRs can account for the $^7$Li abundance successfully. However $^6$Li abundance is only lifted an order of magnitude, which is still much lower than the observed value. As the deuteron abundance is very sensitive to the spectrum choice of BBNCRs, the allowed parameter space for the spectrum is strictly constrained. We should emphasize that the acceleration mechanism for BBNCRs in the early universe is still an open question. For example, strong turbulent magnetic field is probably the solution to the problem. Whether such a mechanism can provide the required spectrum deserves further studies.

\end{abstract}

\newpage
\tableofcontents
\newpage
\section{Introduction}

Big Bang cosmology is an excellent model to describe our Universe \cite{Cosmos}. Albeit its success there is still puzzling ``Lithium problem'' \cite{0409383,0510636} in its building block -- big bang nucleosynthesis (BBN). Namely the theoretical predicted primordial $^7$Li abundance is higher than observation while the $^6$Li abundance is lower. Primordial $^7$Li abundance from measurements of metal-poor halo stars is $^7$Li/H$=(1\sim2)\times10^{-10}$ \cite{0409383}-\cite{1107.1117}, while prediction by the standard BBN (SBBN) model is three to five times higher: $^7$Li/H$=(5.24^{+0.71}_{-0.67})\times10^{-10}$ \cite{0808.2818}. Meanwhile $^6$Li abundance from observation is $^6$Li/H$\approx6\times10^{-12}$ \cite{0510636}, a factor of about 1000 higher than the SBBN model prediction \cite{1011.6179}.

\indent Extensive investigations have focused on Lithium problem. The discrepancy could be due to astrophysical origin \cite{0409672,0608201,1002.1004}. While it is also possible that the discrepancy is arising from Physics beyond the SBBN model. Investigations showed that Lithium abundance would change by varying the nucleon effective couplings as well as the mass parameters, namely the neutron lifetime, the neutron-proton mass difference and the deuteron binding energy etc. \cite{0310892,0610733,0705.0696}. These parameters can be modified due to virtual effects of Physics beyond the Standard Model of particle physics. Another possibility of Physics beyond the SBBN could contain new particles/resonances. Such new particles/resonances participate in thermal nuclear reactions during BBN \cite{0605243}-\cite{1012.0435}, as such Lithium abundance will depend on the detail assumptions about the new particles/resonances. Besides these two approaches, non-thermal electromagnetic \cite{0503023} and hadronic \cite{0402344}-\cite{0907.5003} energy injection was also investigated. Non-thermal particles originate from the nuclear reactions before the final state particles are thermalized. The preliminary studies showed that the mono-energetic injection is hardly the origin of Lithium problem. In this paper we will extend, in some sense, such idea to include the non-thermal energy injection from cosmic rays during and/or shortly after the epoch of BBN.

\indent In this paper, we propose a possible solution to the Lithium problem by not going far from the SBBN model, neither due to the astrophysical factors nor by introduction of new particles. In order to destroy $^7$Li \footnote{ Actually we need to destroy $^7$Be. The relic primordial $^7$Li
mainly (about $90\%$) comes from the decay of $^7$Be.}, we include new contributions from non-thermal particles, namely the accelerated SBBN particles. As is known that the non-thermal particles accelerated in astrophysical environment today are called cosmic rays, it is expected there is such cosmic rays even in the early stage of BBN. We expect that some kind of plasma wave or other mechanisms can feed energy to thermal ions \cite{Tsytovich}, provided that there is strong enough turbulence before/during BBN epoch. As a consequence the particle energy spectrum deviates from thermal distribution. Exploring detailed mechanism on how to induce non-thermal component is beyond the scope of this paper and which will be the focus of the further work. However we propose a toy model to support our basic hypothesis in this paper.

\indent The primordial magnetic fields might be created at some early stage of the evolution of the Universe, for example inflation, the electro-weak phase transition, quark-hadron phase transition and so on. As investigated by authors in ref. \cite{Brandenburg}, after electro-weak phase transition the magnetic field build up and evolve with the expanding Universe. We can estimate the strength of induced electric field through $E\approx\Delta B/\Delta t\times L\approx B~H~L$, where $B\propto R^{-2}$ is the characteristic magnetic field, $H$ is the expansion rate as inverse of the characteristic time, and $L$ is the characteristic length of turbulence which is given by equations (47) and (49) of \cite{Brandenburg}
(here $L=l_0(R/R_0)$ which includes the effect of cosmic expansion). Using the same initial parameters as in ref. \cite{Brandenburg} and extrapolating to $0.01 \mathrm{MeV}$, we get the induced electric field $E\approx 30$ V/m in plasma with temperature $0.01 \mathrm{MeV}$. A charged particle like proton undergoing such electric field will gain energy $\Delta E\approx\tau_{\mathrm{ther}}~v_\parallel~q~E$, where $v_\parallel$ is the particle velocity parallel to the electric field, $q$ is the electric charge of the particle, and $\tau_{\mathrm{ther}}(T)$ is the thermalization time of a high energy (of order MeV) nuclear in plasma of temperature $T$. For O(MeV) protons in plasma of temperature $0.01 \mathrm{MeV}$, $\tau_{\mathrm{ther}}(0.01)\approx 10^{-2}$ s \cite{thermal}, the possible energy gain of an accelerated thermal proton is of order 0.1 MeV. During one free time, a proton can travel a distance of $L_{\mathrm{free}}\approx10^{-3}c$ $\times 10^{-2}$s $\approx 3\times 10^{5}$cm ($c$ is the light velocity) which is just several percent of the characteristic length scale. Protons have the opportunity to be accelerated several times and
thus to gain a couple of MeV energy, though the probability is low. Thus we can expect the energy spectrum of high energy particles will not be too hard.

\indent This proposed toy model needs hydro-magnetic turbulence as the premise. We don't know exactly how hydro-magnetic turbulence develops during the BBN. However according to ref. \cite{Son}, helical hydromagnetic turbulence survives until 100 eV. We found that the correlation length and magnetic field will be larger if we adopt model in ref. \cite{Son} instead. Therefore there should be a considerable parameter space to support our hypothesis.

\indent In this paper we will examine phenomenologically whether such scenario can account for Lithium problem. These non-thermal particles are dubbed as BBN cosmic rays (BBNCRs). In the SBBN conventional abundance calculation, the exothermic reactions are included. BBNCRs contributions to the exothermic SBBN reactions will be investigated. Most importantly, we have to include the endothermic reactions in order to destroy $^7$Be by BBNCRs. In the SBBN, the endothermic reactions are thought not important.

\indent This paper proceeds as following. In section \ref{BBNCRs and Reactions} we describe the required characteristics of BBNCRs, namely the particle species, the energy range and energy spectrum and so on. There are many reactions involving BBNCRs so we need to select which reactions may be important, and detailed discussions on these are contained in this section. The necessary formulae for the abundance computation by including BBNCRs contributions are depicted in section \ref{Computation Scheme and Results}. The numerical results are given in this section. Section \ref{Discussion} contains our conclusions and discussions. Some calculation details are given in Appendix \ref{Classification of Exothermic Reactions} and \ref{Cross Sections and Details of Computation Scheme}.

\section{BBNCRs and Reactions} \label{BBNCRs and Reactions}

In order to estimate the effects of BBNCRs, we need to know the flux, the energy range and the energy spectrum of BBNCRs. They are maybe in evolution
during its work time with the Universe expanding, so the following characters can be seen as an average. Obviously we don't have any knowledge on BBNCRs, therefore we examine some constraints on BBNCRs.

\indent Candidate for BBNCRs may be protons, neutrons, nuclei and electrons. Neutrons are hard to accelerate for electric neutrality, and electrons are out of consideration in SBBN computation, so we focus on nuclei. Because Helium loses energy more quickly through Coulomb scattering \cite{thermal,energyloss1,energyloss2} due to the higher nuclear electric charge \cite{coulomb}, we exclude $^3$He and $^4$He as the components of BBNCRs. We assume that BBNCRs consist of energetic hydrogens, namely protons, deuterons and tritones. In order not to violate the success of SBBN, the intensity of each kind of BBNCRs must be much lower than that of the corresponding SBBN particles. For simplicity we assume that the fraction of BBNCRs is fixed as one single free parameter $\epsilon$.

\indent Next we examine the energy range of BBNCRs. In order not to change deuterium abundance significantly, we must avoid BBNCRs contributions
 from reaction D$(p,n)$2H. Note that the threshold energy and cross section of this reaction are 3.337 MeV and $\approx10^{-2}$ barn respectively. We
  examined the endothermic reactions with threshold energy \cite{qcalc} below 3.337 MeV. In order to destroy $^7$Be by proton, the most effective
   one is $^7$Be$(p,p\alpha)^3$He with threshold energy E$_{\mathrm{th}}=$1.814 MeV. Other endothermic reactions besides $^7$Be$(p,p\alpha)^3$He
    are summarized in table \ref{endothermic}.
\begin{table}[!htb]
\caption{Endothermic reactions}{by energetic hydrogen with threshold energy below 3.337 MeV. \label{endothermic}}
\centering
\begin{tabular}{lll}
    \toprule
    Process & Threshold/MeV & Effect\\
    \midrule
    $^7$Be$(d,^3\mathrm{He})^6$Li & 0.144 & destroy $^7$Be indeed, but less important\\
    & & than $^7$Be$(p,p\alpha)^3$He, for D is much less\\
    & & than $p$; produce $^6$Li\\
    $^7$Li$(d,p)^8$Li & 0.247 & destroy $^7$Li, not important, for we\\
    & & need to destroy $^7$Be\\
    $^7$Be$(d,2p)^7$Li & 0.747 & transformation between $^7$Li and $^7$Be,\\
    & & but less important than $^7$Li$(p,n)^7$Be\\
    T$(p,n)^3$He & 1.019 & transformation between T and $^3$He, and T\\
    & & and $^3$He abundances are assumed to change little\\
    $^7$Li$(d,t)^6$Li & 1.278 & destroy $^7$Li; produce $^6$Li\\
    $^6$Li$(p,p\alpha)$D & 1.721 & destroy $^6$Li, negative, for we\\
    & & need to produce $^6$Li\\
    $^7$Be$(p,p\alpha)^3$He & 1.814 & the most important process to\\
    & & destroy $^7$Be\\
    $^6$Li$(t,np)^7$Li & 1.850 & destroy $^6$Li\\
    $^7$Li$(p,n)^7$Be & 1.880 & transformation between $^7$Li \\
    & & and $^7$Be, important\\
    $^6$Li$(d,\alpha)$2D & 1.967 & destroy $^6$Li\\
    $^7$Be$(d,d\alpha)^3$He & 2.041 & destroy $^7$Be, but less important\\
    $^6$Li$(t,d\alpha)$T & 2.213 & destroy $^6$Li\\
    $^7$Be$(t,t\alpha)^3$He & 2.268 & destroy $^7$Be, but less important\\
    $^3$He$(d,2p)$T & 2.436 & transformation between T and $^3$He\\
    $^7$Be$(d,n)^8$B & 2.688 & destroy $^7$Be, but less important\\
    $^7$Li$(p,p\alpha)$T & 2.821 & destroy $^7$Li\\
    $^7$Li$(d,d\alpha)$T & 3.175 & destroy $^7$Li\\
    \bottomrule
\end{tabular}
\end{table}

\begin{figure}[!htb]
\centering
\includegraphics[scale=1.5]{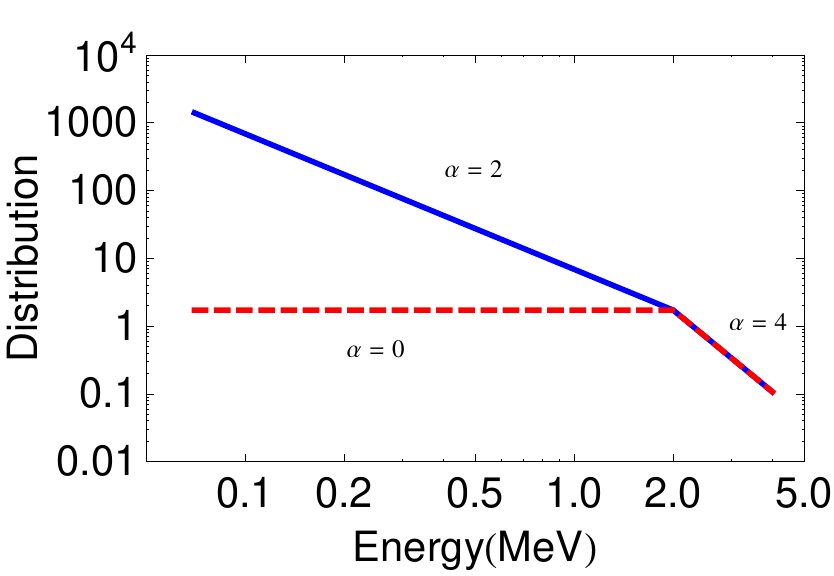}
\caption{BBNCR energy spectrum.}{$\alpha$ is the power index (see eq. (\ref{alpha}) for detail). \label{spectrum}}
\end{figure}

\indent Now we switch to examine the energy spectrum of BBNCRs. In principle the BBNCR energy spectrum will be determined by acceleration mechanism
 and we are lack of knowledge on this. In the realistic case, the energy of BBNCRs should exceed 1.8 MeV but the amount is required to decrease
  quickly, especially above 3.337 MeV, namely the energy threshold of D$(p,n)$2H. Motivated by the power law of cosmic rays today, we assume the
   spectrum of BBNCRs obeys a power law with index 4 from 2 MeV to 4 MeV \footnote{ The energy range is so narrow that small variation of the power
    index will not change the results too much. It is obvious that the larger the power index is, the less important the higher energy particles
     are.}. Below 2 MeV \footnote{ Exactly it is the energy range between 0.09 MeV and 2 MeV. In fact, BBNCRs come from SBBN particles, and the
      lower limit of the energy range of BBNCRs is near or a little above the temperature of SBBN particles. In the SBBN, $^7$Be is produced mainly
       during the epoch when the temperature is between 0.08 MeV and 0.04 MeV, so it seems reasonable that we take 0.09 MeV as the lower limit. Note
        that $^7$Be abundance is not sensitive to the change of the lower limit.}, we choose the power index to be 2 or even in uniform distribution
         like white noise. Our choice of energy spectrum of BBNCRs is depicted in figure \ref{spectrum}.

\indent BBNCRs can affect element abundance via endothermic processes. Besides these BBNCRs will also involve into exothermic SBBN reactions.
 The exothermic processes are thought much less important than endothermic ones in the SBBN relatively. However via exothermic processes the
  non-thermal BBNCRs may have important influence upon the element abundances, especially through which D abundance may be decreased. Complete
   classification of exothermic processes is summarized in Appendix \ref{Classification of Exothermic Reactions}. Reactions added in our numerical
    computation are as following:
\begin{enumerate}
\item Endothermic reactions, including the most important process to destroy $^7$Be -- $^7$Be$(p,p\alpha)^3$He and a weighty origin of $^7$Be
 -- $^7$Li$(p,n)^7$Be;
\item Exothermic reactions, including the collisions between isotopes of hydrogen, especially destroying D -- D$(p,\gamma)^3$He, D$(d,p)$T,
 D$(d,n)^3$He, and T$(d,n)^4$He, and the ones producing or destroying $^7$Li by energetic hydrogen among the most relevant reactions for BBN
  \cite{Iocco} -- $^4$He$(t,\gamma)^7$Li, and $^7$Li$(p,\alpha)^4$He;
\item The ones to produce and destroy $^6$Li, including exothermic reactions $^4$He$(d,\gamma)^6$Li and $^6$Li$(p,\alpha)^3$He and endothermic
 reactions $^7$Be$(d,^3\mathrm{He})^6$Li and $^7$Li$(d,t)^6$Li. Note that the computation for $^6$Li abundance is neither complete nor accurate
  and we will come back to this point in section \ref{Discussion}.
\end{enumerate}
The above reactions are listed in table \ref{total}:
\begin{table}[!htb]
\caption{Additional reactions added in our element abundance computation. \label{total}}
\centering
\begin{tabular}{lll}
    \toprule
    Process & Threshold/MeV & Cross Section\\
    \midrule
    D$(p,\gamma)^3$He & 0 & \cite{dpghe3}\\
    D$(d,p)$T & 0 & \cite{ddpt}\\
    D$(d,n)^3$He & 0 & \cite{ddnhe3}\\
    T$(d,n)^4$He & 0 & \cite{tdnhe4}\\
    $^4$He$(d,\gamma)^6$Li & 0 & \cite{dagli6}\\
    $^4$He$(t,\gamma)^7$Li & 0 & \cite{tagli7}\\
    $^6$Li$(p,\alpha)^3$He & 0 & \cite{li6pahe3,Fiedler}\\
    $^7$Li$(p,\alpha)^4$He & 0 & \cite{Fiedler,li7paa}\\
    $^7$Li$(p,n)^7$Be & 1.880 & \cite{li7pn}\\
    $^7$Be$(p,p\alpha)^3$He & 1.814 & \cite{be7p}\\
    $^7$Li$(d,t)^6$Li & 1.278 & \cite{li7d}\\
    $^7$Be$(d,^3\mathrm{He})^6$Li & 0.144 & \cite{be7d}\\
    \bottomrule
\end{tabular}
\end{table}

\section{Element abundances after including BBNCRs contributions} \label{Computation Scheme and Results}

Now we turn to element abundance computation after including BBNCRs contributions. Besides the standard contributions in SBBN, extra contributions
 come from the processes where the thermal particles collide with non-thermal BBNCRs particles. According to the Boltzmann equation, variation of
  the abundance of nuclide $i$ is described as
\begin{equation}
\frac{dY_i}{dt}=-\sum_jY_iY_j[ij]+\sum_{k,l}Y_kY_l[kl],
\end{equation}
where $[ij]$ is the rate for destroying nuclide $i$ and $[kl]$ is the rate for creating $i$, $Y_i=X_i/A_i$ is the abundance of nuclide $i$ with $X_i$
 the mass fraction and $A_i$ the mass number of nuclide $i$. The sum over $j$ goes through all reactions to destroy nuclide $i$ and the sum over $k,l$
  goes through all reactions to produce nuclide $i$. The rate $[ij]$ is defined by
\begin{equation}
[ij]\equiv N_A\rho\langle ij\rangle,
\end{equation}
where $N_A$ is the Avogadro's number, $\rho$ is the baryon energy density, and $\langle ij\rangle\equiv{\langle\sigma v\rangle}_{ij}$. Here $\sigma$
 is the cross section of the reaction ($ij\to kl$), and $v$ is the relative velocity between these two particles $i$ and $j$. The $\langle\rangle$
  stands for the mean over different relative velocities \cite{stellar}. In the SBBN, $\langle\rangle$ means the thermal average, while in the case
   with BBNCRs it can be computed as
\begin{multline}
\langle\sigma v\rangle_{12}(T,\alpha)=\frac{1}{K_3}\int_{-1}^1\mathrm{d}\cos\theta\frac{1}{K_1}\times\int_{-\infty}^{+\infty} f_1(E_1,T)
\mathrm{d}E_1\times\\\frac{1}{K_2}\int_{0.09}^4 f_2(E_2,\alpha)\mathrm{d}E_2\ \sigma(E_i)v(E_1,E_2,\cos\theta).
\label{eq3}
\end{multline}
Here distribution of SBBN particles $f_1(E_1,T)$ is the normalized Boltzmann distribution,
\begin{equation}
f_1(E_1,T)=2\sqrt{\frac{E_1}{\pi(kT)^3}}e^{-E_1/kT},
\end{equation}
where $k$ is the Boltzmann's constant, $T$ is the Universe temperature, and $E_1$ is the energy of the SBBN particle. Thus the normalization constant
 $K_1=1$ and the energy range is $(-\infty,+\infty)$.
Distribution of BBNCRs is power law with index $\alpha$ (see figure \ref{spectrum}),
\begin{equation}
f_2(E_2,\alpha)\propto E_2^{-\alpha},
\label{alpha}
\end{equation}
\begin{equation}
\alpha=\begin{cases}
2\;\mathrm{or}\;0&\mathrm{for}\quad0.09\,\mathrm{MeV}<E_2<2\,\mathrm{MeV}\\
4&\mathrm{for}\quad2\,\mathrm{MeV}<E_2<4\,\mathrm{MeV}
\end{cases},
\end{equation}
where $E_2$ is the energy of the BBNCR particle, with $\alpha=2$ for power law case and $\alpha=0$ for uniform distribution case. We normalize the
 function $f_2$ from 2 MeV to 4 MeV,
\begin{equation}
K_2=\int_2^4f_2(E_2,\alpha)\mathrm{d}E_2,
\end{equation}
and make $f_2$ continuous at 2 MeV point. When the mass of the SBBN particle is noted as $m_1$ and that of the BBNCR as $m_2$, we can compute the
 relative velocity with the angle of incidence $\theta$,
\begin{equation}
\begin{split}
v=&|\vec{v}_1-\vec{v}_2|\\
=&\sqrt{\frac{2E_1}{m_1}+\frac{2E_2}{m_2}-4\sqrt{\frac{E_1E_2}{m_1m_2}}\cos\theta},
\end{split}
\end{equation}
and incident energy $E_i$
\begin{equation}
E_i=\frac{1}{2}m_iv^2,
\end{equation}
where $m_i$ is the mass of the incident particle. The normalization constant over $\theta$ is $K_3=\int_{-1}^1\mathrm{d}\cos\theta=2$. The cross
 sections in eq. (\ref{eq3}) are taken from experimental data source DataBase EXFOR \cite{exfor} and ENDF\cite{endf}. As a result, the incident
  particle is not simply the energetic particle, but on the criterion from nuclear experiments. Taking $^4$He$(t,\gamma)^7$Li for example, data
   of cross sections are obtained under the situation where $^4$He is taken as the incident particle.

\indent We calculate abundances utilizing the updated version \cite{kawano88,kawano92} of the Wagoner code \cite{wagoner} from \cite{code} with
 appropriate modification in order to include new contributions from BBNCRs. The baryon-to-photon ratio is updated to 6.23$\times10^{-10}$\cite{yita}.
  The rates of endothermal and exothermic processes in table \ref{total} where BBNCRs particles involve are multiplied by $\epsilon$. The new
   contributions are added to the code as new channels. Details of the cross sections we adopted can be found in Appendix
    \ref{Cross Sections and Details of Computation Scheme}.

\indent As the cross section of $^7$Be$(p,p\alpha)^3$He is unknown from experiments, we take D$(p,n)$2H with a shift of the threshold energy as
 a substitute. If the difference is considerable, e.g., the D$(p,n)$2H cross section is $x$ times the $^7$Be$(p,p\alpha)^3$He cross section, we
  need to replace $\epsilon$ as $x\epsilon$.\\

\begin{figure}[!htb]
\centering
\includegraphics{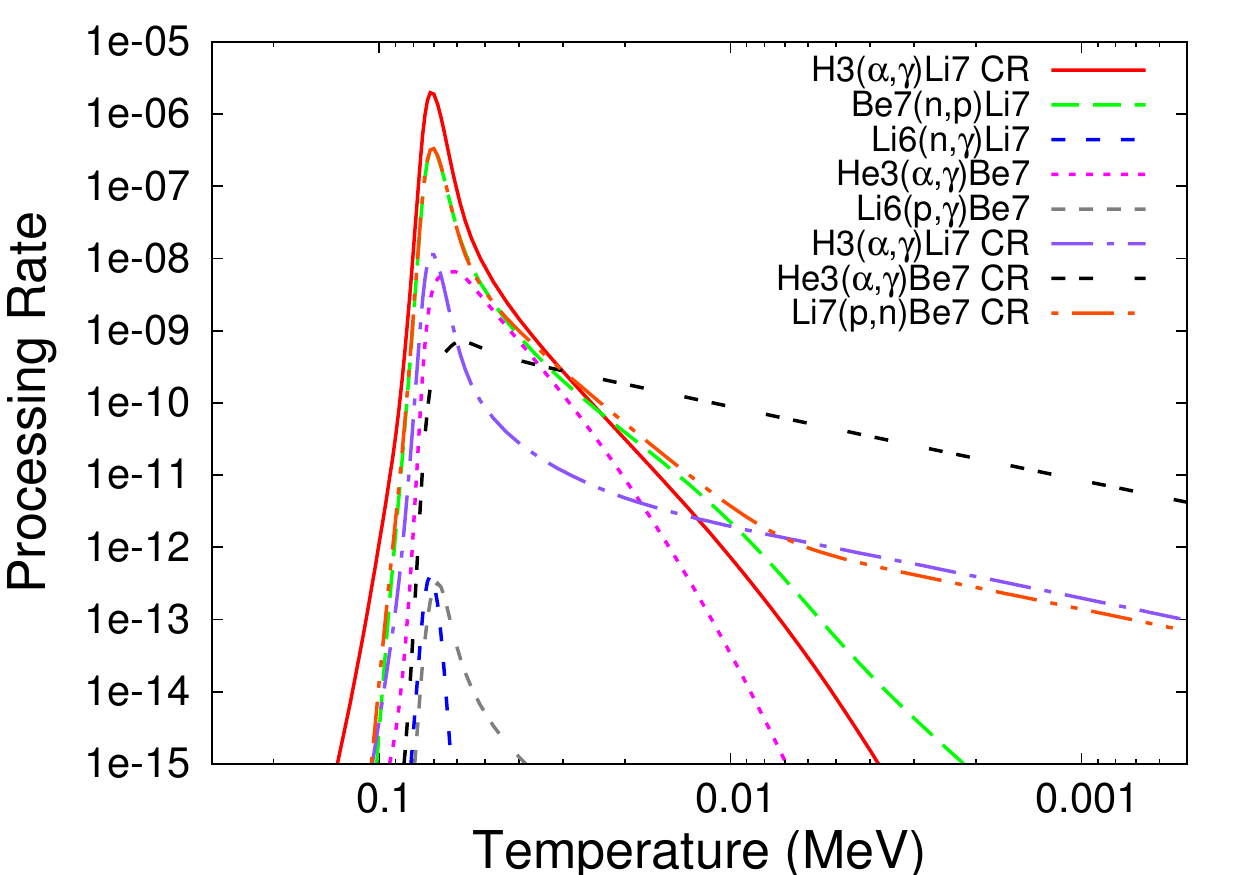}
\caption{Processing rate of producing $^7$Li/$^7$Be as a function of temperature. \label{Li7creatproc}}
\end{figure}

\begin{figure}
\centering
\includegraphics{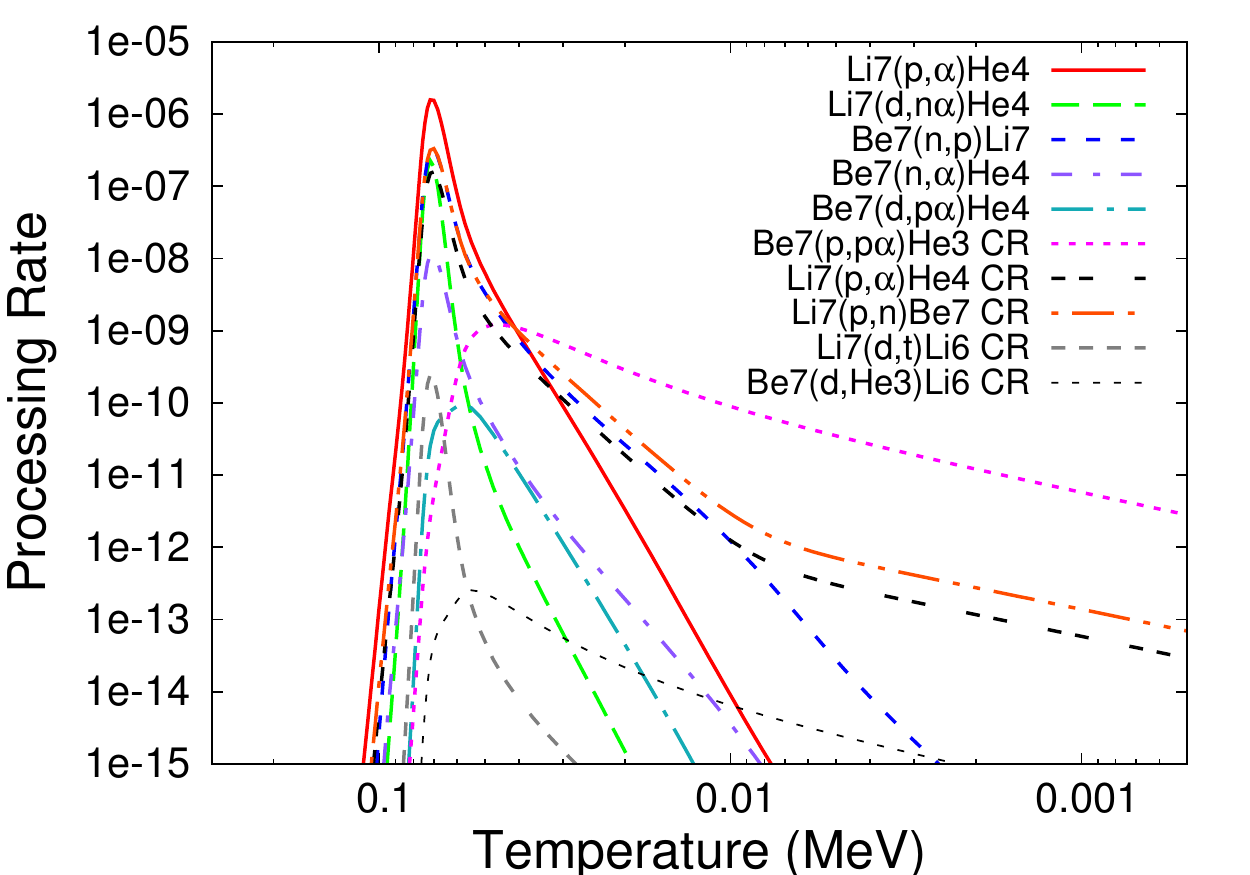}
\caption{Processing rate of destroying $^7$Li/$^7$Be as a function of temperature. \label{Li7destroyproc}}
\end{figure}

\indent Now we turn to show our numerical results for element abundance after including new contributions from BBNCRs. Figure \ref{Li7creatproc}
 and \ref{Li7destroyproc} show processing rates, which is defined as $\epsilon Y_iY_j[ij]/H$\cite{processingrate} ($H$ is the expansion rate),
  of producing and destroying $^7$Li/$^7$Be. From the figures we can clearly see contributions from non-thermal processes, especially in the low
   temperature regime.

\begin{figure}[!htb]
\centering
\includegraphics{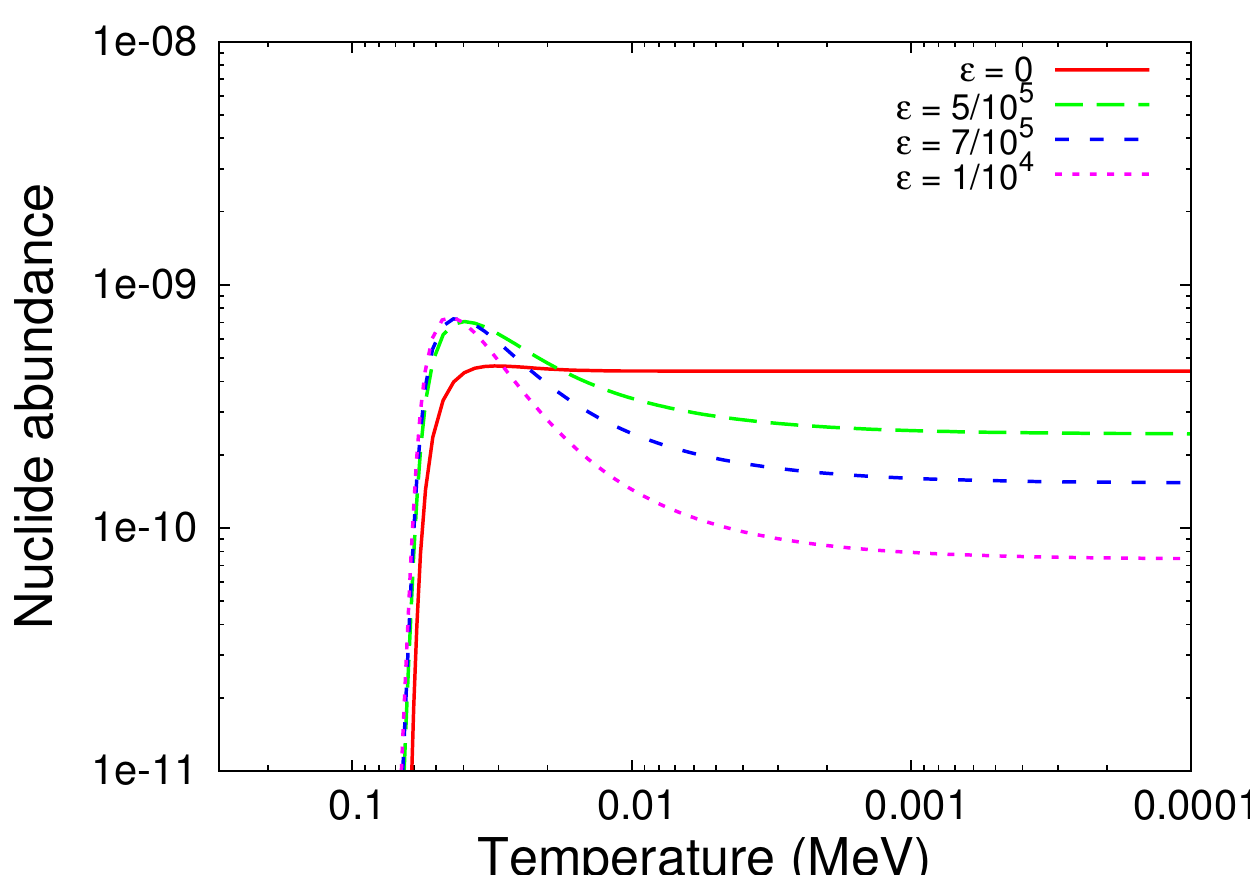}
\caption{$^7$Be abundance as a function of temperature.}{Here $\epsilon$ is taken as 0, $5\times10^{-5}$, $7\times10^{-5}$ and $1\times10^{-4}$
respectively. \label{abBe}}
\end{figure}

\indent $^7$Be abundances under $\alpha=2$ for $0.09\,\mathrm{MeV}<E_2<2\,\mathrm{MeV}$ and different values of $\epsilon$ are shown in figure \ref{abBe},
 among which the red solid line stands for SBBN result ($\epsilon=0$). We can see that $\epsilon=7\times10^{-5}$ can destroy $70\%$ $^7$Be.
  As a price, $5\%$ D is destroyed (as a comparison, the number shifts to $1\%$ for uniform distribution). From the figure we can see clearly
   that new contributions from BBNCRs can account for $^7$Li abundance quite satisfactory. Note that for the processes whose reactants are both
    hydrogen, the effect on D abundance need to be doubled according to the symmetry.

\begin{figure}[!htb]
\centering
\includegraphics{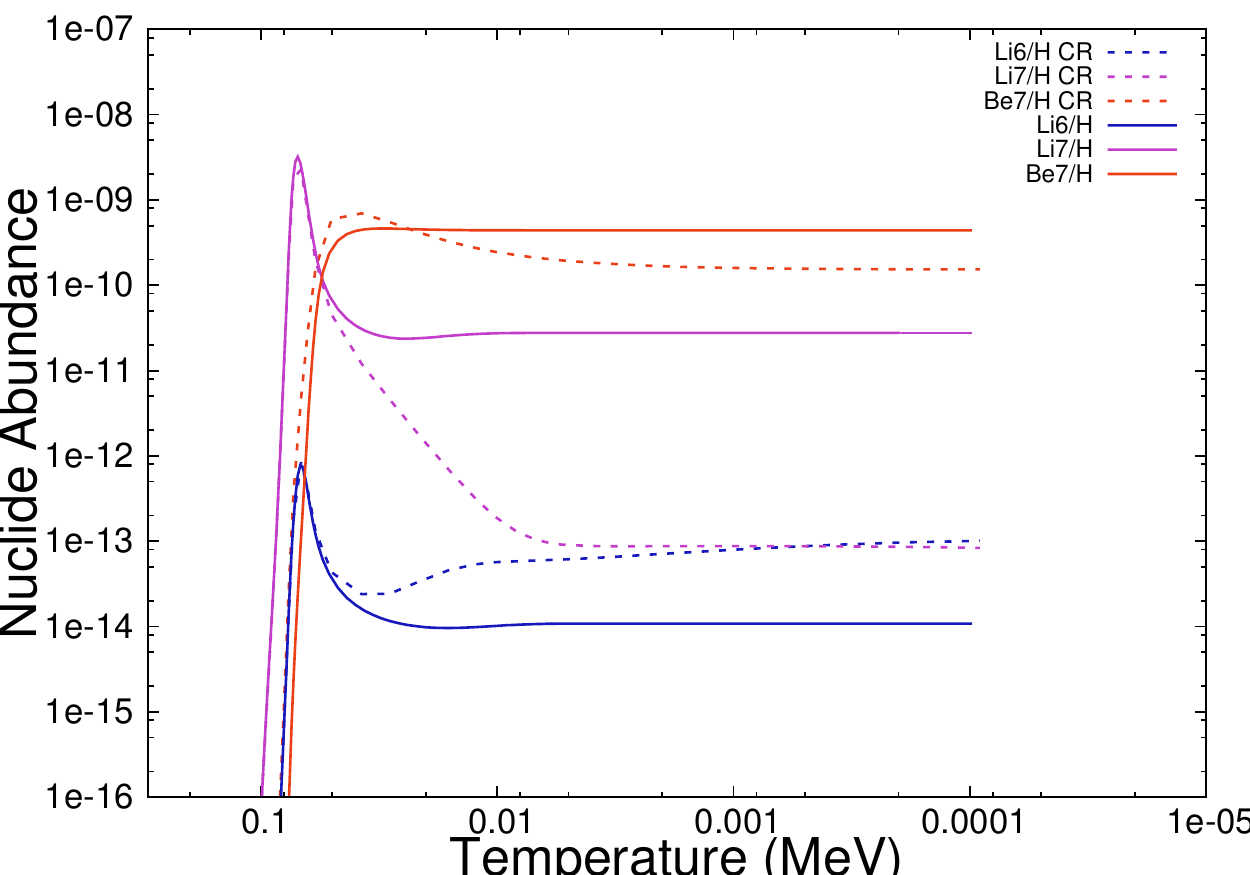}
\caption{Abundances of $^7$Li/$^7$Be and $^6$Li}{as a function of temperature with $\epsilon=7\times 10^{-5}$. \label{abli6li7}}
\end{figure}

\begin{figure}[!htb]
\centering
\includegraphics{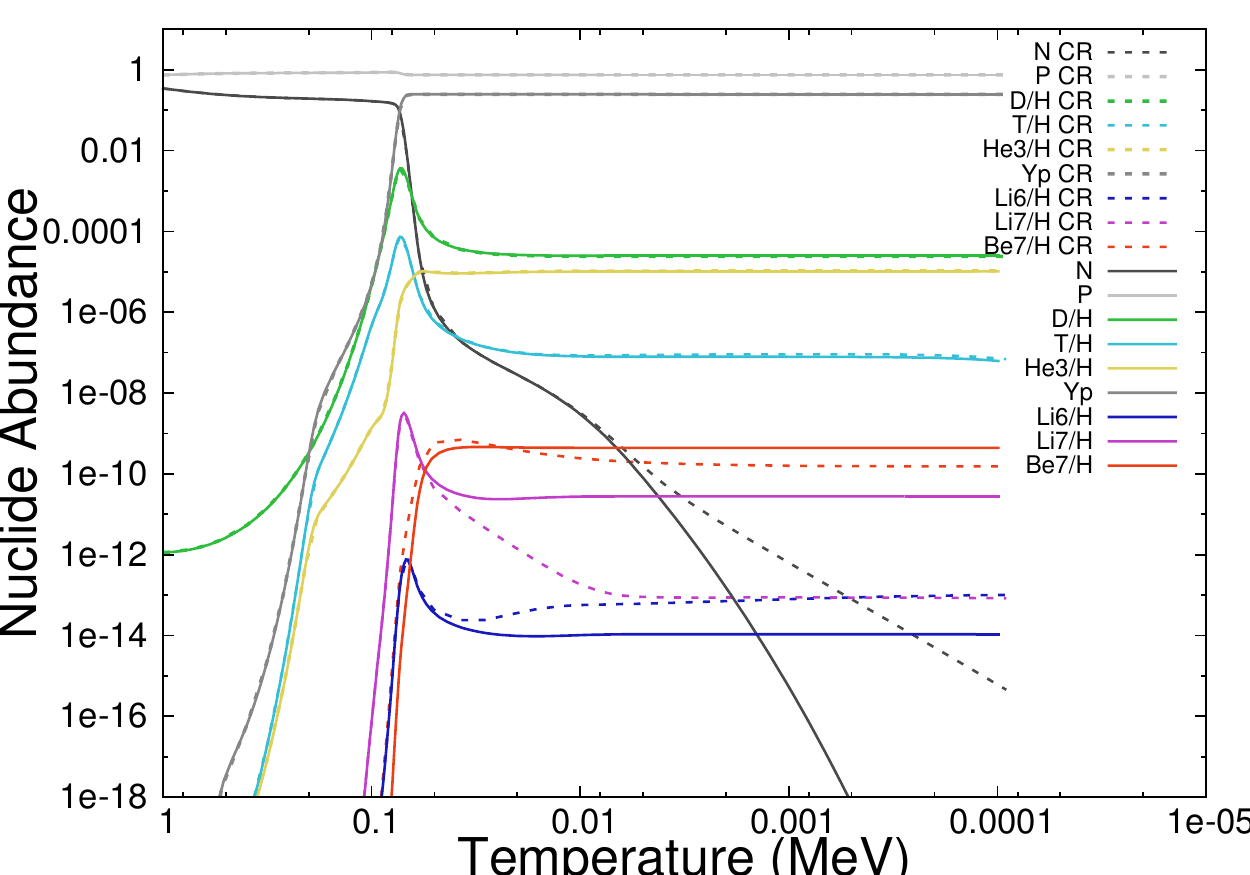}
\caption{Abundances of all elements}{as a function of temperature with $\epsilon=7\times 10^{-5}$. \label{aball}}
\end{figure}

\indent In figure \ref{abli6li7} and \ref{aball} shows abundances of $^7$Li/$^7$Be and $^6$Li, as well as all elements as a function of temperature
 respectively. Here the solid lines stand for the SBBN results, and the dashed lines for results with BBNCRs under $\alpha=2$ for
  $0.09\,\mathrm{MeV}<E_2<2\,\mathrm{MeV}$ and $\epsilon=7\times10^{-5}$. While the BBNCRs can account for $^7$Li abundance, for the $^6$Li abundance,
   energetic BBNCRs can enhance $^6$Li one order of magnitude, but not enough compared to observations. From the curves we can also see that the
    shifts of other element abundances after including BBNCRs contributions are usually tiny.

\begin{figure}[!htb]
\centering
\includegraphics{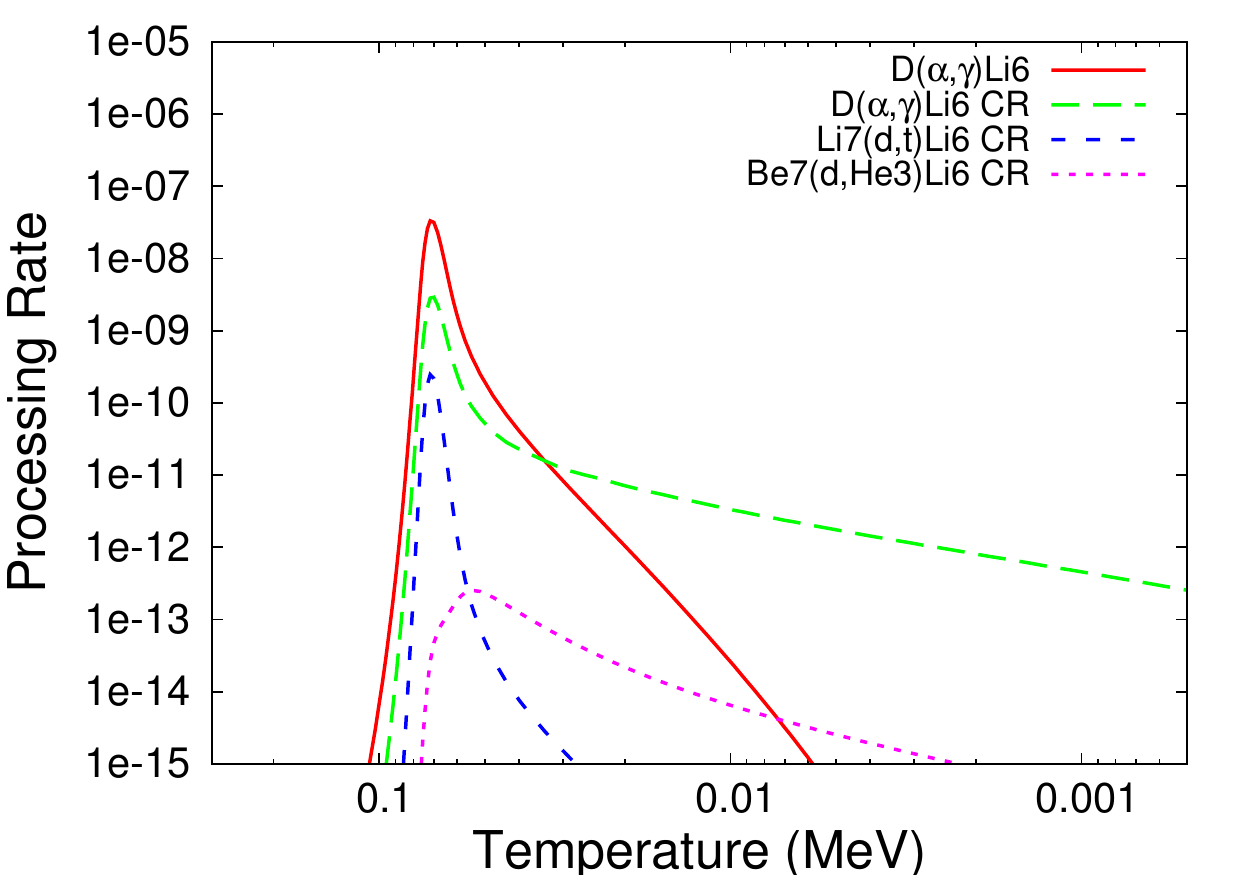}
\caption{Processing rate of producing $^6$Li as a function of temperature. \label{Li6creatproc}}
\end{figure}

\begin{figure}
\centering
\includegraphics{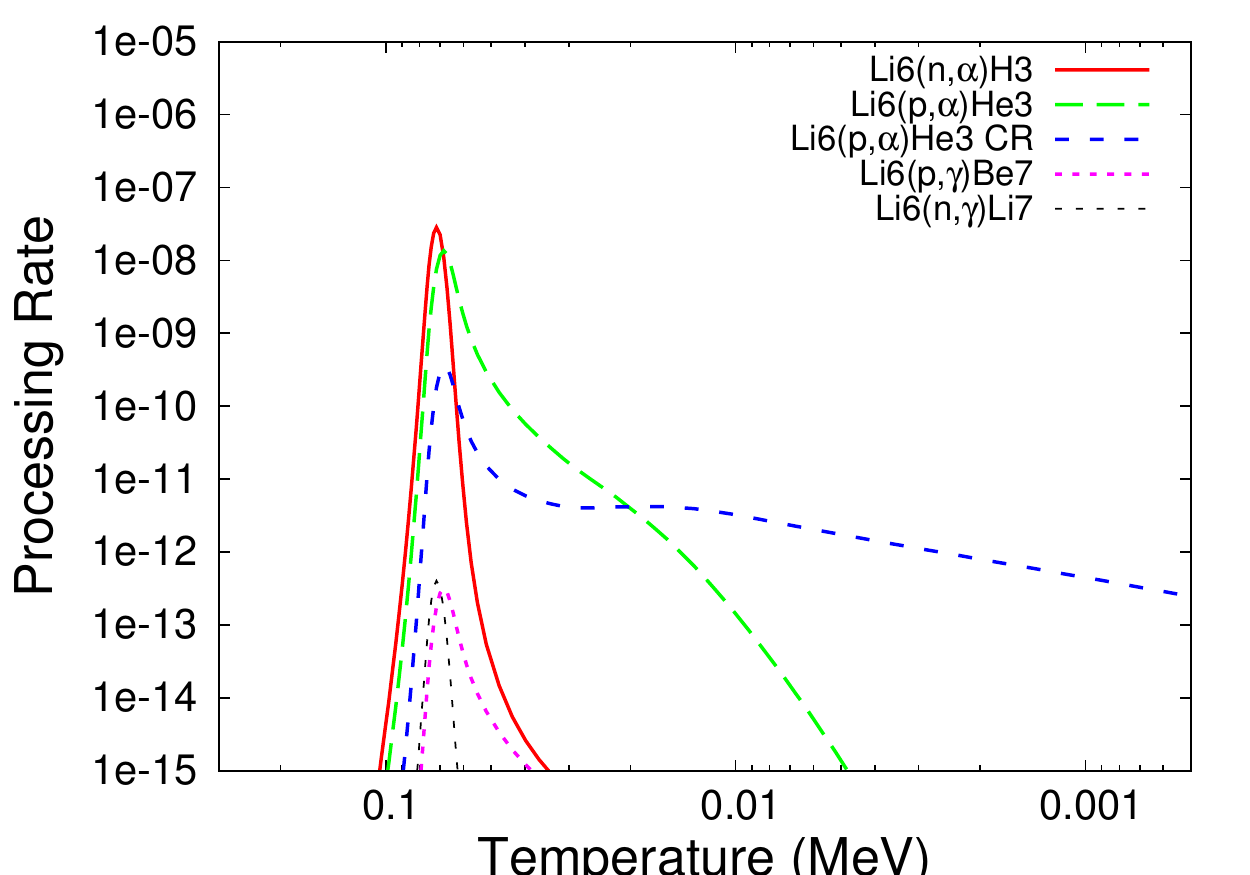}
\caption{Processing rate of destroying $^6$Li as a function of temperature. \label{Li6destroyproc}}
\end{figure}

\indent For completeness we also show processing rates of producing and destroying $^6$Li in figure \ref{Li6creatproc} and \ref{Li6destroyproc}.
 It is not hard to see that non-thermal contributions are quite similar to the $^7$Li case. BBNCRs play important role mainly when the
  Universe temperature falls below 0.04 MeV.

\section{Conclusions and discussions} \label{Discussion}

In this paper, we investigated whether cosmic rays in the BBN epoch (BBNCRs) can account for Lithium problem or not. In order to keep the success
 of SBBN, the flux, energy range and spectrum of BBNCRs are severely constrained. In the allowed parameter space, extra contributions from BBNCRs
  to $^7$Li abundance can fill the discrepancy between SBBN prediction and observations. However BBNCRs can lift $^6$Li abundance in the SBBN an
   order of magnitude, but still less than that of observations.

\indent We need to point out some factors beyond investigations in the paper. First, measurements of cross sections of nuclear reactions,
 especially $^7$Be$(p,p\alpha)^3$He is critical. Lack of knowledge on T$(t,\gamma)^6$He ($^6$He will decay to $^6$Li in 0.8s) and $^3$He$(t,\gamma)^6$Li
  cross sections brings uncertainty on $^6$Li results, too. If their cross section between 2 MeV and 4 MeV is about $O$(mb), the contributions
   to $^6$Li abundance is comparable to that of $^4$He$(d,\gamma)^6$Li.

\indent Second, we consider hydrogen as the only component of BBNCRs in this paper, however the possible $^3$He or $^4$He as BBNCRs particles will
 change our results. We can imagine that $^3$He and $^4$He are accelerated to the same energy and spectrum (in fact their energy should be lower and
  the spectrum be softer than hydrogen). In this case, effects of some reactions (e.g. $^4$He$(d,\gamma)^6$Li and $^4$He$(t,\gamma)^7$Li) will be
   doubled. At the same time, new reactions are triggered by energetic $^3$He and $^4$He (see table \ref{he3he4}). The reaction
    $^7$Be$(\alpha,p)^{10}$B will destroy $^7$Be and produce $^{10}$B.

\begin{table}[!htb]
\belowcaptionskip=0.4cm
\caption{List of new reactions triggered by energetic $^3$He and $^4$He. \label{he3he4}}
\centering
\begin{tabular}{lll}
    \toprule
    Process & Threshold/MeV\\
    \midrule
    $^6$Li$(^3\mathrm{He},2p)^7$Li & 0.703\\
    $^7$Li$(^3\mathrm{He},t)^7$Be & 1.259\\
    $^7$Be$(\alpha,p)^{10}$B & 1.800\\
    $^6$Li$(^3\mathrm{He},d\alpha)^3$He & 2.213\\
    $^7$Be$(^3\mathrm{He},\alpha)2^3$He & 2.268\\
    $^6$Li$(\alpha,d)2^4$He & 2.455\\
    $^7$Be$(\alpha,2\alpha)^3$He & 2.491\\
    $^6$Li$(\alpha,d)^8$Be & 2.608\\
    $^7$Be$(\alpha,^3\mathrm{He})^8$Be & 2.635\\
    $^6$Li$(^3\mathrm{He},d)^8$B & 2.965\\
    $^6$Li$(^3\mathrm{He},np)^7$Be & 3.171\\
    \bottomrule
\end{tabular}
\end{table}

\begin{figure}
\belowcaptionskip=0.4cm
\centering
\includegraphics[scale=1.5]{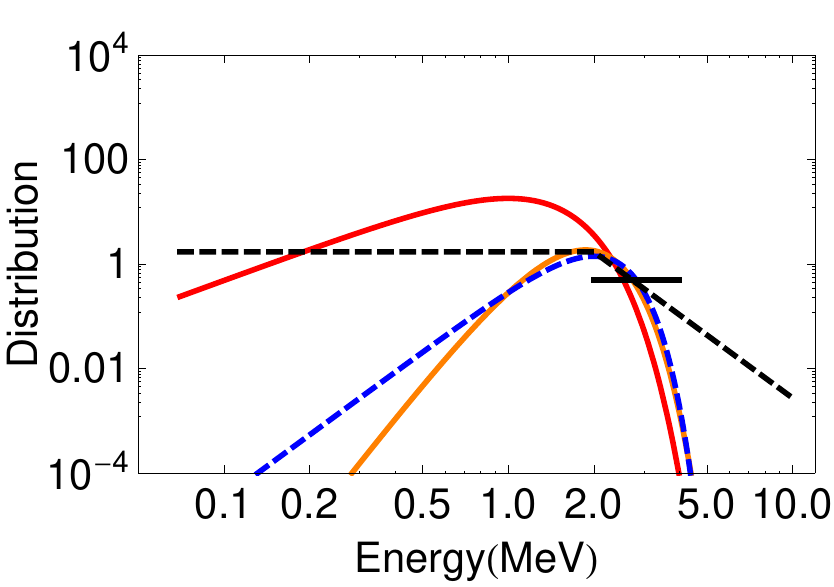}
\caption{Energy spectrum test. \label{energytest}}
\end{figure}

\indent In our analysis, it is important that the amount of BBNCRs should be very low above 3.337 MeV, namely the D$(p,n)$2H threshold. In order to test the effect of the spectrum choice on the deuteron abundance, we tried different spectrums, and show some representative ones in figure \ref{energytest}. For each spectrum, we give the necessary fraction of BBNCRs, namely $\epsilon$, to destroy 70\% $^7$Be, and the corresponding percentage of destructed deuterium. The results are shown in table \ref{3.337}. The upper limit of BBNCRs spectrum is taken to be 10 MeV \footnote{ We ignore endothermic reactions whose threshold is higher}. We find that BBNCRs above 3.337 MeV would destroy deuterium inevitably. Therefore, the main part of the realistic spectrum should lie between 1.814 MeV and 3.337 MeV, and a rapid drop below 1.814 MeV and above 3.337 MeV is favorable.

\begin{table}[!htb]
\belowcaptionskip=0.4cm
\caption{The effect of the spectrum choice on the deuteron abundance. \label{3.337}}
\scriptsize
\centering
\begin{tabular}{llll}
    \toprule
    Description of spectrum & Curves in figure \ref{energytest} & $\epsilon$ fraction & D destroyed\\
    \midrule
    uniform distribution between 2 MeV and 4 MeV & black, solid & 2$\times10^{-5}$ & 1.9\%\\\hline
    uniform distribution between 0.09 MeV and 2 MeV, & & &\\
    $E^{-4}$ between 2 MeV and 10 MeV & black, dashed & 3$\times10^{-5}$ & 10.6\%\\\hline
    $E^2e^{-E^2}$ between 0.09 MeV and 10 MeV & red, solid & 3$\times10^{-4}$ & 15\%\\\hline
    $E^7e^{-E^2}$ between 0.09 MeV and 10 MeV & orange, solid & 1.5$\times10^{-4}$ & 1.8\%\\\hline
    $E^4e^{-(E/1.814)^3}$ between 0.09 MeV and 10 MeV & blue, dashed & 1$\times10^{-4}$ & 1.2\%\\
    \bottomrule
\end{tabular}
\end{table}

\indent At last but not least, the critic issue for the BBNCRs is about acceleration mechanism during BBN and whether the required flux and energy
 range as well as spectrum can be induced. Anyhow the energetic particles are thought to be easily thermalized by photons and nuclei around before
  they collide with other nuclei \cite{energyloss1,energyloss2}.

\section*{Acknowledgment}

We would like to thank Xiao-jun Bi and Jie Meng for the stimulating discussions. This work was supported in part by the Natural Science Foundation
 of China (Nos. 10725524, 11075003, 11135003 and 1135010) and Chinese Academy of Sciences (No. KJCX2-YW-N13).

\appendix

\section{Classification of Exothermic Reactions \label{Classification of Exothermic Reactions}}

Exothermic reactions where BBNCRs involve are listed in two tables. Table \ref{exothermicall1} shows SBBN reactions which are included in the Wagoner code and table \ref{exothermicall2} shows reactions besides those.
\begin{longtable}{ll}
\caption{Exothermic reactions -- Part \uppercase\expandafter{\romannumeral1} \label{exothermicall1}}\\
    \toprule
    Process & Effect\\
    \midrule
    \endfirsthead
    \midrule
    Process & Effect\\
    \midrule
    \endhead    \midrule
    \multicolumn{2}{r}{To be continued\dots}\\
    \endfoot
    \bottomrule
    \endlastfoot
    H$(n,\gamma)$D & produce D, but not important, for $n$ is too little\\
    D$(n,\gamma)$T & destroy D, but not important, for $n$ is too little\\
    D$(p,\gamma)^3$He & destroy D, see table \ref{total}\\
    D$(d,n)^3$He & destroy D, see table \ref{total}\\
    D$(d,p)$T & destroy D, see table \ref{total}\\
    T$(p,\gamma)^4$He & destroy T\footnote{ T will decay to $^3$He, and observations only set the upper limit of $^3$He\cite{he3}.}\\
    T$(d,n)^4$He & destroy D, see table \ref{total}; destroy T, see the footnote\\
    $^3$He$(d,p)^4$He & destroy D, but less important than D$(p,\gamma)^3$He;\\
     & destroy $^3$He, see the footnote\\
    $^4$He$(d,\gamma)^6$Li & produce $^6$Li, see table \ref{total}\\
    $^4$He$(t,\gamma)^7$Li & produce $^7$Li, see table \ref{total}\\
    $^6$Li$(p,\gamma)^7$Be & destroy $^6$Li, but less important than $^6$Li$(p,\alpha)^3$He, for the\\
    & electromagnetic cross section is smaller than the strong one\\
    $^6$Li$(p,\alpha)^3$He & destroy $^6$Li, see table \ref{total}\\
    $^7$Li$(p,\alpha)^4$He & destroy $^7$Li, see table \ref{total}\\
    $^7$Li$(d,n)2^4$He & destroy $^7$Li, but less important than $^7$Li$(p,\alpha)^4$He\\
    $^7$Be$(d,p)2^4$He & destroy $^7$Be, but less important\\
    $^7$Be$(p,\gamma)^8$B & destroy $^7$Be, but less important, for the electromagnetic\\
    & cross section\\
\end{longtable}

\begin{longtable}{ll}
\caption{\label{exothermicall2}Exothermic reactions -- Part \uppercase\expandafter{\romannumeral2}}\\
    \toprule
    Process & Effect\\
    \midrule
    \endfirsthead
    \midrule
    Process & Effect\\
    \midrule
    \endhead
    \midrule
    \multicolumn{2}{r}{To be continued\dots}\\
    \endfoot
    \bottomrule
    \endlastfoot
    D$(d,\gamma)^4$He & destroy D, but less important than D$(d,p)$T or\\
    & D$(d,n)^3$He, for the electromagnetic cross section\\
    T$(t,2n)^4$He & destroy T, but change little compared to the SBBN situation\\
    T$(t,\gamma)^6$He & produce $^6$Li, see Section \ref{Discussion}\\
    $^3$He$(t,\gamma)^6$Li & produce $^6$Li, see Section \ref{Discussion}\\
    $^3$He$(t,d)^4$He & produce D, but change little compared to the SBBN situation\\
    $^3$He$(t,np)^4$He & destroy T and $^3$He, but change little compared to the SBBN situation\\
    $^6$Li$(d,\alpha)^4$He & destroy $^6$Li, but less important\\
    $^6$Li$(d,p)^7$Li & destroy $^6$Li, but less important\\
    $^6$Li$(d,n)^7$Be & destroy $^6$Li, but less important\\
    $^6$Li$(d,pt)^4$He & destroy $^6$Li, but less important\\
    $^6$Li$(d,n^3\mathrm{He})^4$He & destroy $^6$Li, but less important\\
    $^6$Li$(t,\gamma)^9$Be & destroy $^6$Li, but less important;\\
    & produce $^9$Be, but less important than $^7$Li$(t,n)^9$Be,\\
    & for the electromagnetic cross section\\
    $^6$Li$(t,n)2^4$He & destroy $^6$Li, but less important\\
    $^6$Li$(t,d)^7$Li & destroy $^6$Li, but less important\\
    $^6$Li$(t,p)^8$Li & destroy $^6$Li, but less important\\
    $^7$Li$(d,\gamma)^9$Be & produce $^9$Be, but less important than $^7$Li$(t,n)^9$Be,\\
    & for the electromagnetic cross section\\
    $^7$Li$(t,\gamma)^{10}$Be & produce $^{10}$B, ($^{10}$Be decays to $^{10}$B)\\
    & but less important than $^7$Be$(\alpha,p)^{10}$B\\
    $^7$Li$(t,n)^9$Be & produce $^9$Be, maybe important, constraints from observations\cite{Be9}\\
    $^7$Li$(t,\alpha)^6$He & produce $^6$Li, but less important than T$(t,\gamma)^6$He or $^3$He$(t,\gamma)^6$Li\\
    $^7$Li$(t,2n)2^4$He & destroy $^7$Li, but less important\\
    $^7$Be$(t,\gamma)^{10}$B & produce $^{10}$B, but less important than $^7$Be$(\alpha,p)^{10}$B\\
    $^7$Be$(t,\alpha)^6$Li & produce $^6$Li, but less important than T$(t,\gamma)^6$He or $^3$He$(t,\gamma)^6$Li\\
    $^7$Be$(t,d)2^4$He & destroy $^7$Be, but less important\\
    $^7$Be$(t,p)^9$Be & produce $^9$Be, maybe important, constraints from observations\cite{Be9}\\
    $^7$Be$(t,np)2^4$He & destroy $^7$Be, but less important\\
    $^7$Be$(t,^3\mathrm{He})^7$Li & transformation between $^7$Li and $^7$Be, but less important than\\
    & $^7$Li$(p,n)^7$Be\\
\end{longtable}

\section{Cross Sections and Details of Computation Scheme \label{Cross Sections and Details of Computation Scheme}}

\begin{figure}[!htb]
\begin{center}
\scalebox{0.455}{\includegraphics[bb=0 0 504 336]{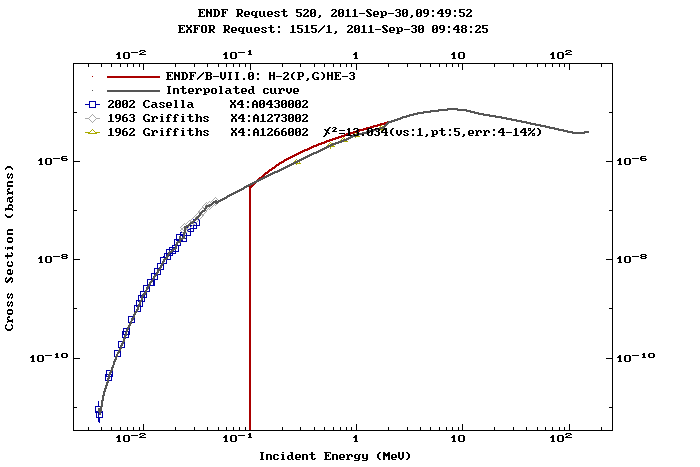}}
\caption{Cross sections of D$(p,\gamma)^3$He \label{dpgh3}}
\end{center}
\end{figure}

\begin{figure}[!htb]
\begin{center}
\scalebox{0.455}{\includegraphics[bb=0 0 504 336]{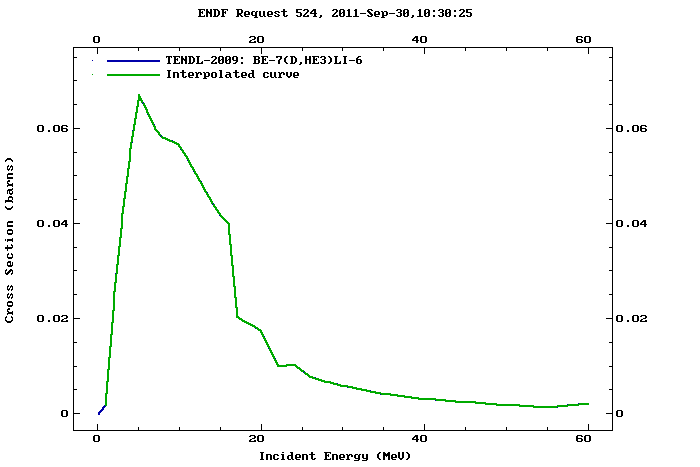}}
\caption{Cross sections of $^7$Be$(d,^3\mathrm{He})^6$Li \label{be7dhe3li6}}
\end{center}
\end{figure}

\indent Cross sections of the reactions included in our computation are shown in figure \ref{dpgh3}-\ref{be7dhe3li6}. The possible errors here can be
 accommodated by changing the amount of BBNCRs, namely varying the free parameter $\epsilon$.

\indent Cross sections below the first available point from the experiment is set to zero, and above the last available point the cross sections are
 set to be flat. Among the experimental points, cross sections are interpolated simply linearly. However for D$(p,\gamma)^3$He, $^7$Li$(p,\alpha)^4$He,
  $^6$Li$(p,\alpha)^3$He and T$(d,n)^4$He, the cross sections are interpolated double-logarithmic linearly, in order to make them more smooth.

\indent The integral limit over thermal spectrum is not infinite. Instead we use $0.001T$ as the lower limit and $8 T$ as the upper limit, where $T$
 is the Universe temperature. Integration is carried out in six-point Gaussian scheme.

\newpage

\begin{figure}[!htb]
\begin{center}
\scalebox{0.455}{\includegraphics[bb=0 0 504 336]{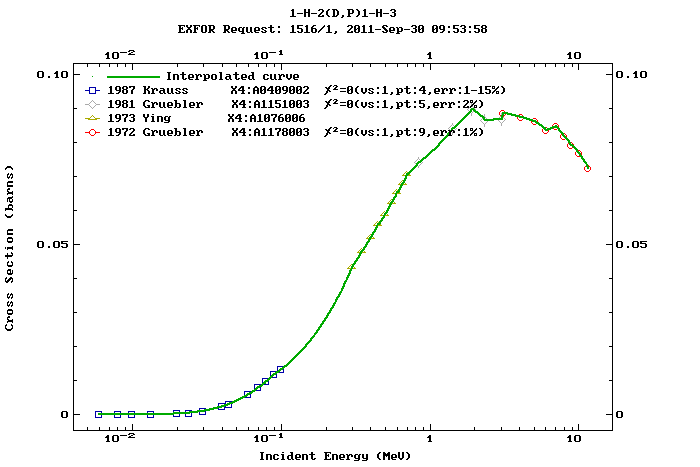}}
\caption{Cross sections of D$(d,p)$T}
\end{center}
\end{figure}

\begin{figure}[!htb]
\begin{center}
\scalebox{0.455}{\includegraphics[bb=0 0 504 336]{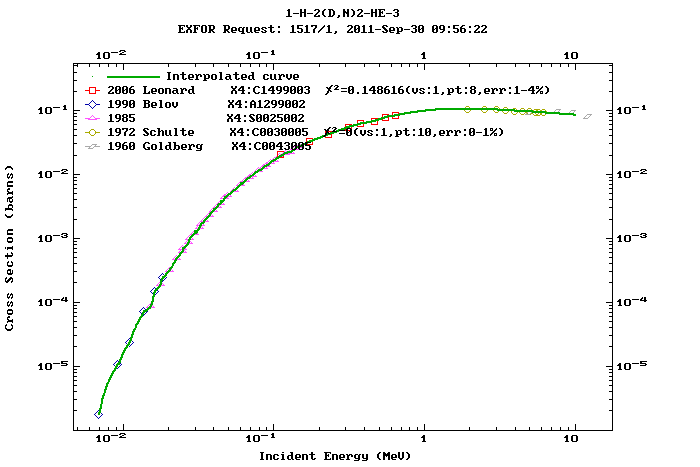}}
\caption{Cross sections of D$(d,n)^3$He}
\end{center}
\end{figure}

\begin{figure}[!htb]
\begin{center}
\scalebox{0.455}{\includegraphics[bb=0 0 504 336]{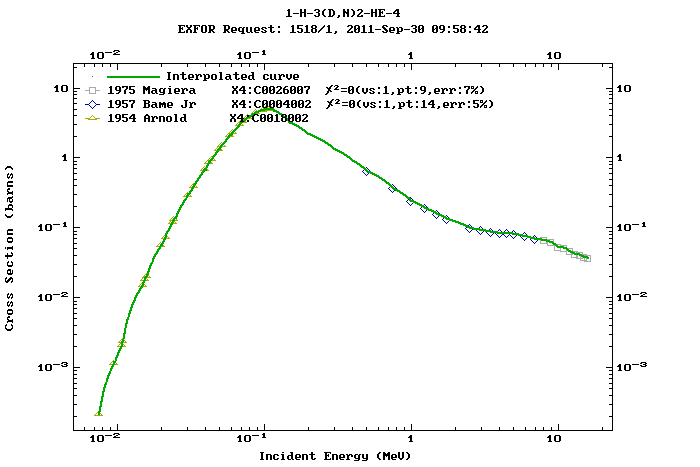}}
\caption{Cross sections of T$(d,n)^4$He}
\end{center}
\end{figure}

\begin{figure}[!htb]
\begin{center}
\scalebox{0.455}{\includegraphics[bb=0 0 504 336]{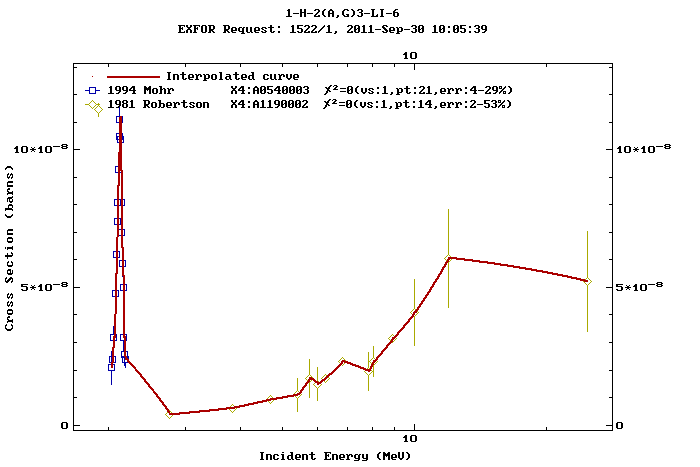}}
\caption{Cross sections of $^4$He$(d,\gamma)^6$Li}
\end{center}
\end{figure}

\begin{figure}[!htb]
\begin{center}
\scalebox{0.455}{\includegraphics[bb=0 0 504 336]{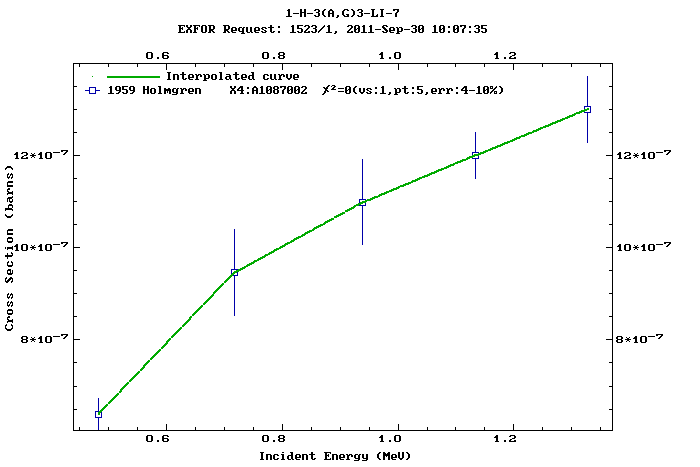}}
\caption{Cross sections of $^4$He$(t,\gamma)^7$Li}
\end{center}
\end{figure}

\begin{figure}[!htb]
\begin{center}
\scalebox{0.455}{\includegraphics[bb=0 0 504 336]{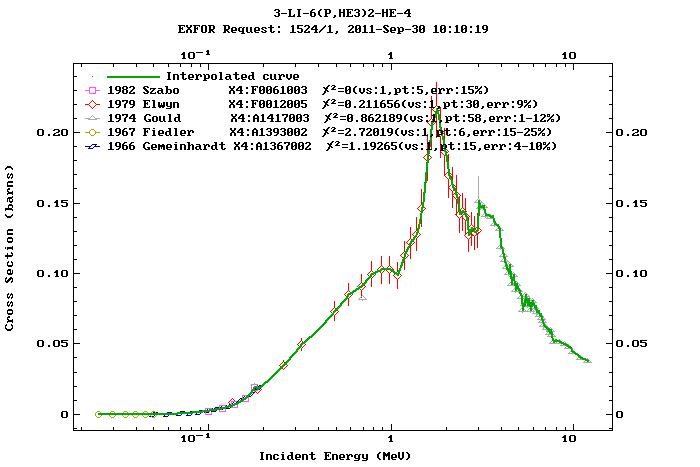}}
\caption{Cross sections of $^6$Li$(p,\alpha)^3$He}
\end{center}
\end{figure}

\begin{figure}[!htb]
\begin{center}
\scalebox{0.455}{\includegraphics[bb=0 0 504 336]{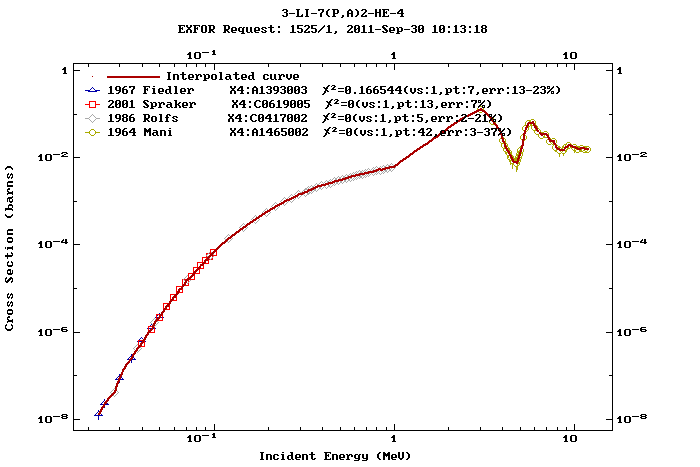}}
\caption{Cross sections of $^7$Li$(p,\alpha)^4$He}
\end{center}
\end{figure}

\begin{figure}[!htb]
\begin{center}
\scalebox{0.455}{\includegraphics[bb=0 0 504 336]{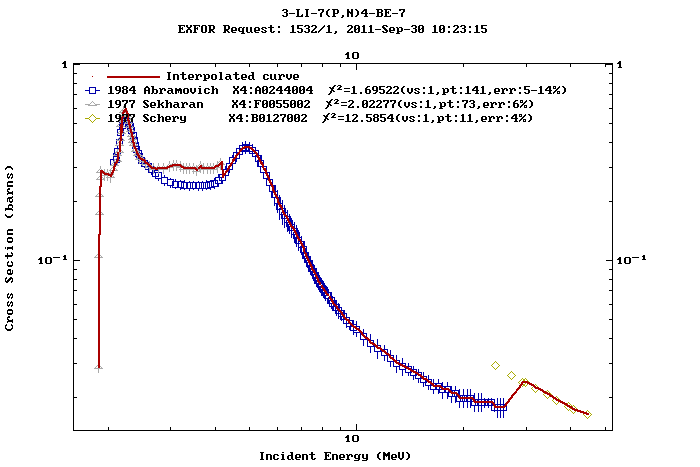}}
\caption{Cross sections of $^7$Li$(p,n)^7$Be}
\end{center}
\end{figure}

\begin{figure}[!htb]
\begin{center}
\scalebox{0.455}{\includegraphics[bb=0 0 504 336]{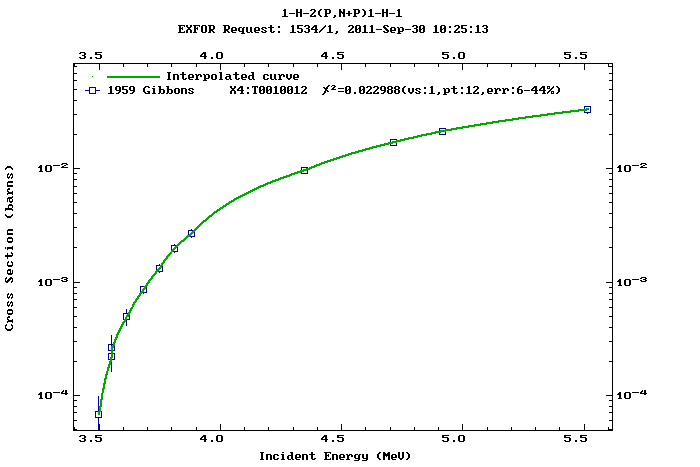}}
\caption{Cross sections of D$(p,n)$2H}
\end{center}
\end{figure}

\begin{figure}[!htb]
\begin{center}
\scalebox{0.455}{\includegraphics[bb=0 0 504 336]{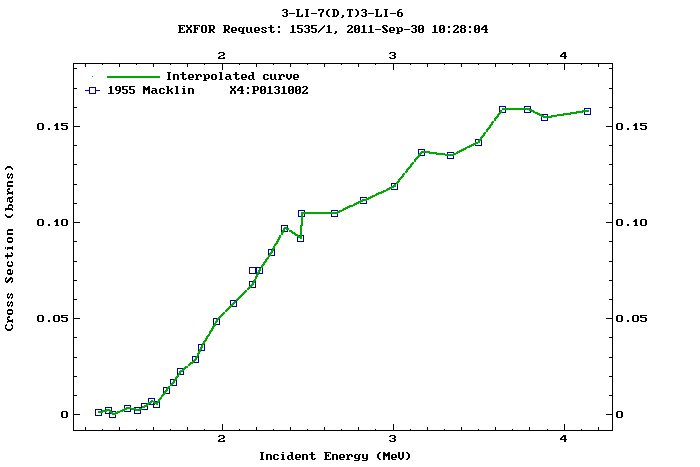}}
\caption{Cross sections of $^7$Li$(d,t)^6$Li}
\end{center}
\end{figure}

\end{document}